\documentclass[sigconf]{acmart}

\usepackage[utf8]{inputenc}
\usepackage[T1]{fontenc} 

\usepackage{natbib}

\usepackage{algorithmic}
\usepackage{algorithm}
\usepackage{array}
\usepackage[caption=false,font=normalsize,labelfont=sf,textfont=sf]{subfig}
\usepackage{textcomp}
\usepackage{stfloats}
\usepackage{url}
\usepackage{verbatim}
\usepackage{graphicx}

\usepackage{hyperref}
\usepackage{csquotes}
\usepackage{enumitem}

\usepackage{colortbl}
\usepackage{xcolor}
\usepackage{multirow}
\usepackage{tabularx}
\usepackage{booktabs}
\usepackage[super]{nth}

\newcommand*{\enq}[1]{\enquote{{\itshape#1}}}

\copyrightyear{2026}
\acmYear{2026}
\setcopyright{cc}
\setcctype{by}
\acmConference[MSR '26]{23rd International Conference on Mining Software Repositories}{April 13--14, 2026}{Rio de Janeiro, Brazil}
\acmBooktitle{23rd International Conference on Mining Software Repositories (MSR '26), April 13--14, 2026, Rio de Janeiro, Brazil}
\acmDOI{10.1145/3793302.3793350}
\acmISBN{979-8-4007-2474-9/2026/04}

\begin{document}
\begin{sloppy}

\title{Context Engineering for AI Agents in Open-Source Software}

\author{Seyedmoein Mohsenimofidi}
\affiliation{%
  \institution{Heidelberg University}
  \country{Germany}
}
\email{s.mohsenimofidi@uni-heidelberg.de}
\orcid{0009-0009-1620-2735}

\author{Matthias Galster}
\affiliation{%
 \institution{University of Bamberg}
 \country{Germany}
}
\email{mgalster@ieee.org}
\orcid{0000-0003-3491-1833}

\author{Christoph Treude}
\affiliation{%
  \institution{Singapore Management University}
  \country{Singapore}
}
\email{ctreude@smu.edu.sg}
\orcid{0000-0002-6919-2149}

\author{Sebastian Baltes}
\affiliation{%
  \institution{Heidelberg University}
  \country{Germany}
}
\email{sebastian.baltes@uni-heidelberg.de}
\orcid{0000-0002-2442-7522}

\renewcommand{\shortauthors}{Mohsenimofidi et al.}

\begin{abstract}
GenAI-based coding assistants have disrupted software development.
The next generation of these tools is agent-based, operating with more autonomy and potentially without human oversight.
Like human developers, AI agents require contextual information to develop solutions that are in line with the standards, policies, and workflows of the software projects they operate in.  
Vendors of popular agentic tools (e.g., Claude Code) recommend maintaining version-controlled Markdown files that describe aspects such as the project structure, code style, or building and testing. 
The content of these files is then automatically added to each prompt.
Recently, \texttt{AGENTS.md} has emerged as a potential standard that consolidates existing tool-specific formats.
However, little is known about whether and how developers adopt this format.
Therefore, in this paper, we present the results of a preliminary study investigating the adoption of AI context files in 466 open-source software projects.
We analyze the information that developers provide in \texttt{AGENTS.md} files, how they present that information, and how the files evolve over time.
Our findings indicate that there is no established content structure yet and that there is a lot of variation in terms of how context is provided (descriptive, prescriptive, prohibitive, explanatory, conditional). 
Our commit-level analysis provides first insights into the evolution of the provided context. 
AI context files provide a unique opportunity to study real-world context engineering. 
In particular, we see great potential in studying which structural or presentational modifications can positively affect the quality of the generated content.
\end{abstract}




\maketitle

\section{Introduction}

The launches of GitHub Copilot in 2021 and ChatGPT in 2022 have started to transform how software is developed.
Today, generative AI (GenAI) tools built around large language models (LLMs) support software engineers throughout the software development lifecycle (SDLC)---although most published work focuses on code and test generation~\cite{DBLP:journals/tosem/HouZLYWLLLGW24, DBLP:conf/fose-ws/FanGHLSYZ23, DBLP:journals/ese/ZhengNZCCGWW25, DBLP:journals/tse/WangHCLWW24, DBLP:journals/tse/TufanoDMCB24, 10.1145/3747588}.
The \emph{Devin AI} demo published in March 2024 fueled the first hype around agent-based software development~\cite{knightDevin}, but it took until 2025 for agent-based software development to reach considerable adoption.
In February 2025, Anthropic released \emph{Claude Code}, a \enq{command line tool for agentic coding}~\cite{anthropicClaudeCodeRelease}, which represents a further step toward more autonomous AI-assisted software development, enabling developers to assign coding tasks to AI agents via a terminal interface.
Human oversight is still built-in, but can be turned off by the developer.
This brings GenAI assistants closer to the inherent meaning of an \emph{agent} that operates autonomously, adapts to change, and creates and pursues goals (from the Latin `agere', `to do' in English)~\cite{russell2021}.

\emph{Context engineering} is the deliberate process of designing, structuring, and providing task-relevant information to LLMs~\cite{DBLP:journals/corr/abs-2507-13334, philschmidSkillPrompting}.
While \emph{prompt engineering} focuses on how a task is described to the model (e.g., instructions and output indicators)~\cite{DBLP:journals/corr/abs-2402-07927, promptingguideElementsPrompt}, context engineering focuses on what task-relevant information the model has access to, including relevant guidelines, configuration files, documentation, and exemplary code snippets~\cite{DBLP:journals/corr/abs-2507-13334, philschmidSkillPrompting}.
An advantage of agent-based tools compared to conversational tools is that they allow persistent, structured, and task-specific context to be provided in a more fine-grained and targeted manner~\cite{dexterHorthyComplexCodebases}.
One way to ``engineer'' context is to add machine-readable AI context files to source code repositories.
The AI agents then automatically add the content of these files to their prompts.
While traditional \texttt{README} files are written for humans, AI context files are explicitly designed for AI agents, providing a central machine-readable source of contextual information.
Their content can include everything from the required terminal commands to build and test the project over documentation links, common workflows, coding conventions, to instructions for creating pull requests.

\texttt{AGENTS.md} was introduced as an open tool-agnostic convention for such AI context files~\cite{agentsmd}, and it was recently announced as a project in the \emph{Agentic AI Foundation}. 
OpenAI's \emph{Codex} tool relies on this format \cite{openaiIntroducingCodex}, while \emph{Claude Code} by default searches for a file named \texttt{CLAUDE.md}.
Anthropic's best-practice guide recommends teams to put that file into version control so that all team members benefit from consistent AI behavior~\cite{anthropicClaudeCode}.
GitHub introduced a similar AI context file for \emph{Copilot} named \texttt{copilot-instructions.md}~\cite{githubCopilotCode}.

Since prompts are only rarely preserved after content has been generated \cite{DBLP:conf/msr/TafreshipourIHA25}, AI context files offer a unique opportunity to study how developers customize AI agents to their needs, what information they consider relevant to include, how they present it, and how instructions and contextual descriptions evolve.
To our knowledge, we present the first holistic empirical study that analyzes context files used to guide different AI agents in open-source software (OSS) projects.
We addressed the following research questions:

\begin{description}[style=multiline, labelindent=4mm, leftmargin=12mm, topsep=4pt]
\item[RQ1] \emph{How widely have OSS projects adopted AI context files?} 
\item[RQ2] \emph{What information do open-source developers provide in AI context files and how do they present it?}
\item[RQ3] \emph{How do AI context files evolve over time?}
\end{description}

An initial search in October 2025 suggested that tens of thousands of GitHub repositories already contained AI context files~\cite{gitHubSearchAgents}.
However, there was no systematic analysis focusing on ``engineered'' software projects.
This paper provides a first step toward filling this gap by mining GitHub repositories to study how AI context files are adopted, structured, and maintained, with the overarching goal of understanding how software teams engage in context engineering in practice.
The results of our preliminary study are based on data collected from 10,000 GitHub repositories (\textbf{RQ1}). For \textbf{RQ2} and \textbf{RQ3}, we performed a detailed qualitative analysis of relevant repository data, including file content, commits, and issues.
As the first study that explores the above aspects, we follow an exploratory bottom-up approach to build an initial understanding of AI context files as novel software artifacts. 
We will extend our study in the future to answer the research questions more holistically.

\section{Related Work}


Agentic GenAI tools promise to introduce autonomous decision-making and proactive problem-solving along the SDLC~\cite{DBLP:journals/jcis/HughesDMSAJDACDFSJKKLMS25, DBLP:conf/kbse/SuriDSDSK23}.
Advances in LLMs, reinforcement learning, and multi-agent frameworks enabled the implementation of software agents that go beyond simple prompt-response interactions~\cite{DBLP:journals/chinaf/XiCGHDHZWJZZFWXZWJZLYDW25}. 
One of the first agent-based software development tools was \emph{Devin AI}~\cite{cognitionCognitionIntroducing}, which allowed agents to search the web, edit files, and execute commands to complete tasks iteratively and independently.
In the academic community, \emph{SWE-agent}~\cite{DBLP:conf/nips/YangJWLYNP24} allowed LLM-based agents to communicate with the repository environment by reading, modifying, and executing bash commands~\cite{DBLP:conf/nips/YangJWLYNP24}.
Another example is \emph{AutoCodeRover}~\cite{DBLP:conf/issta/0002RFR24}, which enabled agents to access code search APIs to help them find methods within specific classes for bug location identification~\cite{DBLP:conf/issta/0002RFR24}.
\citeauthor{DBLP:conf/kbse/SuriDSDSK23} showcased the potential of autonomous agents, particularly \emph{Auto-GPT}, in software engineering tasks and demonstrates the importance of context-specific prompts in complex frameworks~\cite{DBLP:conf/kbse/SuriDSDSK23}.
However, they also revealed that \emph{Auto-GPT}~\cite{agptAutoGPT}, despite performing well on simpler tasks, struggles with ambiguity and complexity, underscoring the need for accurate context.
Although context plays a critical role in guiding autonomous agents, prompts are typically treated as temporary artifacts and are rarely preserved or reused~\cite{DBLP:journals/corr/abs-2509-17548, DBLP:journals/corr/abs-2509-17096, DBLP:journals/corr/abs-2509-12421}.
This lack of prompt management limits reproducibility~\cite{baltes2025guidelinesempiricalstudiessoftware} and underscores the importance of making prompt and context information explicit and manageable by using versioned AI context files.
Recently, researchers have started investigating AI context files as novel software artifacts~\cite{10.1007/978-3-032-12089-2_40}.
However, a holistic analysis across tools and formats has, to the best of our knowledge, not been published yet.

\section{Data Collection}
\label{sec:data_collection}

\begin{figure*}[tb]
    \centering
    \includegraphics[width=1\linewidth]{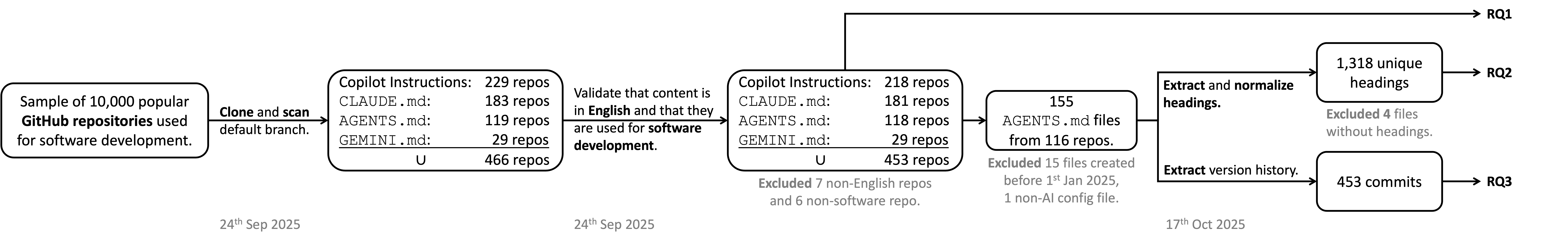}
    \caption{Data collection process.}
    \Description{Diagram showing the data collection process.}
    \label{fig:data-collection}
\end{figure*}

To answer our research questions, we collected AI context files from OSS projects on GitHub.
Since GitHub hosts not only ``engineered' software projects, we needed to develop a strategy for selecting repositories~\cite{DBLP:journals/ese/MunaiahKCN17}.
Many popular repositories on GitHub are not software projects~\cite{gitHubSearchPopularRepos}, which complicates sampling. 

Our starting point for selecting true software projects was the SEART GitHub search tool~\cite{seartGitHubSearch, DBLP:conf/msr/DabicAB21}.
We selected non-fork repositories that have at least two contributors, have a license, and were created before \nth{1} January 2024 with commits since \nth{1} June 2024.
We then excluded archived, disabled, or locked repositories, resulting in a first sample of 228,890 repositories.
In the next step, we selected repositories with an OSI-compliant open-source license~\cite{osiLicenses} and then manually filtered out licenses not intended for software and licenses with low adoption, i.e., used in fewer than 261 repositories (median). 
We focused on the ten most popular languages (Python, TypeScript, JavaScript, Go, Java, C++, Rust, PHP, C\#, and C) and excluded repositories with fewer than 271 commits (median) or fewer than 7 watchers (median).
This resulted in a final sample of 48,795 repositories.
The purpose of this filtering process was to select repositories that represent actively maintained software projects with a sufficiently long development history, written in one of the major programming languages.
For this paper, our goal was to start with mature popular repositories. 
Thus, we selected 10,000 repositories based on a ranking approach that balances popularity (\#stars, \#watchers, \#contributors) and maturity (\#commits to default branch, project age, LOC).

Figure~\ref{fig:data-collection} outlines our data collection process for these 10,000 repositories. 
We cloned them and scanned their default branch to find all types of context files that GitHub Copilot supports: Copilot instructions, \texttt{CLAUDE.md}, \texttt{AGENTS.md}, and \texttt{GEMINI.md}~\cite{gitHubCopilotSupportedFormats}.
We then manually checked all repositories to exclude non-English and non-software projects.
We used the resulting data to answer \textbf{RQ1} (see Section~\ref{sec:rq1}).
Since \texttt{AGENTS.md} is the only format that serves as an open tool-agnostic convention, we decided to focus on it to answer \textbf{RQ2} and \textbf{RQ3} (see Sections~\ref{sec:rq2} and \ref{sec:rq3}).
Our data collection and analysis scripts and the analyzed data are available online~\cite{supplementaryMaterial}.

\section{Results}



\subsection{Adoption (RQ1)}
\label{sec:rq1}

Only 466 (5\%) of the repositories that we scanned had already adopted at least one of the formats we considered, reflecting that we are still in an early stage of adoption.
One limitation is our focus on four selected tools. 
We consider extending the analysis to cover more tools (and more repositories) an important direction for our future work.
It will also be interesting to study trends over time, e.g., whether projects converge toward one file format or whether tool-specific formats persist.

The distribution of languages was roughly aligned with the languages' general representation in our sample, although Go was slightly overrepresented.
We found AI context files in 135 repositories with TypeScript as main language, 58 Go, 58 Python, 56 C\#, 36 Java, 34 JavaScript, 32 C++, 29 Rust, 19 PHP, and 9 C.
Certain file types were more prevalent in certain programming languages.
C\#, for example, had a strong focus on \emph{Copilot}, while \emph{Claude Code} was very popular for TypeScript.
We also investigated which files most commonly co-occurred, finding that (\texttt{AGENTS.md}, \texttt{CLAUDE.md}) was the most common pair (25 repositories). 

\subsection{Information and Structure (RQ2)}
\label{sec:rq2}

Before we dive into our answer to \textbf{RQ2}, we briefly characterize the content of AI context files.
Copilot instruction files were on average the longest ($M=310$ lines, $SD=127$ lines), followed by \texttt{CLAUDE.md} files ($M=287$, $SD=112$); \texttt{GEMINI.md} files were the shortest ($M=106$, $SD=65$).
Interestingly, \texttt{AGENTS.md} files had the highest variation in file length ($M=142$, $SD=231$).
This variation may reflect the amount of information that developers provide.
Exploring whether the context length is proportional to the project size or other factors is part of our future work.

To answer \textbf{RQ2}, we extracted all section headings from the 155 \texttt{AGENTS.md} files in our sample, converted them to lower case, removed special characters, and lemmatized the words to be able to group semantically equivalent variations (e.g., ``tests'' and ``testing'').
We excluded 15 \texttt{AGENTS.md} files that were created before \nth{1} January 2025, i.e., before the \texttt{AGENTS.md} convention was introduced.
For each lemmatized heading, we determined (1) in how many distinct repositories it occurred, (2) in how many distinct files it occurred (\#files > \#repositories), and (3) how many total occurrences it had (multiple equivalent headings per file).
We excluded five repositories that contained \texttt{AGENTS.md} files without any heading structure.
For the remaining files, we recorded the heading levels (from \#, i.e., level 1, to \#\#\#\#\#, i.e., level 5) to understand their structural depth.
Following related work on \texttt{README} files~\cite{DBLP:journals/tse/GaoTZ25}, which found that the first- and second-level headings are the most informative and consistent, we restricted our initial analysis to these levels. 

We manually developed an initial coding guide based on the 44 lemmatized section headings that appeared in $\ge 3$ different repositories and $\ge 3$ times at heading levels 1 or 2.
We then examined examples of section content for each heading.
The resulting coding guide, shown in Table~\ref{tab:rq2_categories}, groups semantically similar headings into categories.
We then applied this guide to a larger set of 91 lemmatized section headings that were used in $\ge 2$ repositories and $\ge 2$ times at heading levels 1 or 2.
This analysis provides an overview of the information most commonly provided in \texttt{AGENTS.md} files.
Topics such as code conventions and best practices, contribution guidelines, and architecture or project structure appeared frequently, in contrast to sections on troubleshooting or security.

We also noticed differences in writing style when analyzing the files.
To examine these differences more closely, we analyzed all 50 sections labeled \textsc{Conventions}, the most common category in our dataset.
We found that the writing style can be characterized along five stylistic dimensions: \emph{descriptive}, \emph{prescriptive}, \emph{prohibitive}, \emph{explanatory}, and \emph{conditional}.
Some sections were \emph{descriptive}, documenting existing conventions without giving explicit instructions, e.g., \enq{This project uses the Linux Kernel Style Guideline.}
Such statements summarize current practices or configurations that the AI agent should be aware of, rather than prescribing behavior.
Others were \emph{prescriptive}, written as direct imperatives that instruct how to act, e.g., \enq{Follow the existing code style and conventions.} 
This style provides explicit behavioral rules and was often formatted as concise bullet points.
\emph{Prohibitive} statements were also common, explicitly indicating what not to do, e.g., \enq{Never commit directly to the main branch.}
These prohibitions set boundaries and clarify the constraints that AI agents should respect. 
Some projects added short explanations after the rules, resulting in an \emph{explanatory} style, e.g., \enq{Avoid hard-coded waits to prevent timing issues in CI environments.}
Here, the justification (\enq{to prevent timing issues}) provides context for why a convention exists. 
Finally, we observed \emph{conditional} formulations that specify what to do in certain situations, e.g., \enq{If you need to use reflection, use \texttt{ReflectionUtils} APIs.}
This style encodes situational logic, specifying conditional actions that depend on the context of the agent's task.

In summary, \texttt{AGENTS.md} files vary widely both in the information they contain and how they are presented, yet some recurring patterns are emerging.
Projects often document architecture, contribution processes, and coding conventions, but without a consistent structure.
Stylistic choices range from \emph{descriptive} to \emph{directive}, reflecting experimentation with how best to communicate expectations to AI agents.
These observations suggest that conventions for documenting context are still evolving and point to promising opportunities for future work on how information structure and style influence agent behavior.

\subsection{Evolution (RQ3)}
\label{sec:rq3}

To answer \textbf{RQ3}, we analyzed the commit histories of all 155 \texttt{AGENTS.md} files and found that 77 (50\%) of them had not been changed, 36 (23\%) only once, and 32 (21\%) between two and seven times.
For this study, we were primarily interested in understanding the types of changes developers make in AI context files.
We decided to focus on the 10 files (6\%) with at least 10 commits, which yielded a sample of 169 commits to annotate (37\% of all collected commits).
The resulting modification patterns varied per file, with some histories spanning a short period with many changes (e.g., \href{https://github.com/neomjs/neo/commits/dev/AGENTS.md}{\texttt{neomjs/neo}}: 49 changes over 19 days) and others spanning longer periods with fewer changes (e.g., \href{https://github.com/gofiber/fiber/commits/main/AGENTS.md}{\texttt{gofiber/fiber}}: 11 changes, 148 days).
Although we did not analyze files with fewer than 10 commits in detail, we noticed that the history of these files varied significantly as well, ranging from 0 to 127 days, with 2 to 8 commits.
A more detailed analysis of evolution patterns is an important direction for future work.


To understand what developers change in AI context files and to inductively identify general change categories, we manually reviewed the commits, including the source code diff, the commit messages, and any related issues or pull requests.
Two authors developed an initial coding guide and then iteratively refined the emerging categories and descriptions. 
We considered each commit as an isolated change.
Although most commits (111 commits, i.e., 66\%) represented only one change category, we observed commits that represented multiple.
Table~\ref{tab:change_categories} shows that some categories 
refer to the overall structure of \texttt{AGENTS.md} files while others 
refer to the content of specific sections, indicating a narrower scope.
Note that we currently do not quantify changes per category, meaning that, e.g., `\textbf{Add} \textsc{section(s)}' can represent a change that added one or multiple sections.
Furthermore, we did not label the intent of a change, e.g., 
why a certain section was added or removed.
An analysis of the intent of changes is the subject of our future work.

\begin{table}[tb]
\centering
\footnotesize
\caption{Categories of information provided in \texttt{AGENTS.md} files (last column: number of level~1 and 2 headings).}
\begin{tabularx}{\linewidth}{p{1.82cm}Xr}
\toprule
\textbf{Category} & \textbf{Description} & \textbf{\#} \\
\midrule
\textsc{Conventions} & Outlines coding standards, naming/formatting conventions, and best practices for writing consistent and maintainable code. & 50 \\

\textsc{Contribution guidelines} & Provides instructions for contributing to the repository, such as branching, code reviews, or CI requirements. & 48 \\

\textsc{Architecture/ structure} & Describes how the project or repository is organized, including key directories, modules, components, and relationships between them. & 47 \\

\textsc{Build commands} & Lists commands for building, running, or deploying. & 40 \\ 

\textsc{Goals/purposes} & Summarizes what the project or agent does, its goals or purposes, and high-level functionality or capabilities. & 32 \\

\textsc{Test execution} & Explains how to execute test suites or individual tests, including tools, commands, and environments. & 32 \\

\textsc{Metadata} & Contains file metadata or configuration (e.g., tags). & 29 \\ 

\textsc{Test strategy} & Describes the overall approach to testing (unit, integration, end-to-end), test organization, or principles guiding test coverage and design. & 24 \\

\textsc{Tech stack} & Lists programming languages, libraries, frameworks, or other dependencies used in the project. & 15 \\

\textsc{Setup} & Covers installation prerequisites, environment setup, and initial steps required to run/use the project locally. & 11 \\

\textsc{References} & Provides a concise list of frequently used commands, API references, or quick tips for developers or users. & 9 \\

\textsc{Troubleshooting} & Offers guidance for diagnosing and resolving common errors, failures, or configuration problems encountered during development or deployment. & 8 \\

\textsc{Patterns/examples} & Shows reusable patterns, sample agent configs, or example use cases to guide understanding or extensions. & 8 \\

\textsc{Security} & Highlights security-related advice, configurations, or precautions (e.g., managing secrets or access controls). & 6 \\ 
\bottomrule
\end{tabularx}
\label{tab:rq2_categories}
\end{table}

\begin{table}[tb]
\centering
\footnotesize
\caption{Categories of changes for AGENTS.md files with $\ge 10$ commits (last column: category frequency across all files).}
\begin{tabularx}{\linewidth}{p{2.5cm}Xr}
\toprule
\textbf{Category} & \textbf{Description} & \textbf{\#} \\
\midrule
\textbf{Add} \textsc{instruction(s)} & Add instruction line(s) to existing sections. & 78\\
\textbf{Modify} \textsc{instruction(s)} & Modify instruction line(s) within a section (ignoring typo fixes and references additions). & 59\\
\textbf{Add} \textsc{section(s)} & Add new section(s) to the \texttt{AGENTS.md} file. & 26\\
\textbf{Remove} \textsc{instruction(s)} & Remove line(s) with instructions from existing sections. & 23\\
\textbf{Modify} \textsc{heading(s)} & Modify existing section heading title or level. & 23\\
\textbf{Modify} \textsc{text} & Minor changes to content of \texttt{AGENTS.md} file, such as fixing typos. & 19\\
\textbf{Reformat} \textsc{style} & Changing visual appearance of content in \texttt{AGENTS.md} file (not related to structure). & 10\\
\textbf{Remove} \textsc{section(s)} & Remove sections from \texttt{AGENTS.md} file. & 2\\
\textbf{Update} \textsc{reference(s)} & Update references, e.g., URLs. & 2\\
\bottomrule
\end{tabularx}
\label{tab:change_categories}
\end{table}

Table~\ref{tab:change_categories} shows that the most frequent change categories are `\textbf{Add} \textsc{instruction(s)}' and `\textbf{Modify} \textsc{instruction(s)}'.
For all examined \texttt{AGENTS.md} files, these categories occurred as the first or second change in the history of changes.
Looking at commit messages related to changes, we found a few interesting cases.
For example, for \texttt{AGENTS.md} in \texttt{rsyslog}, one of the commit messages states \enq{AI support: Agent shall no longer call stylecheck.sh}.
The related change category is `\textbf{Remove} \textsc{instruction(s)}', because the change deleted an instruction.
A commit in \texttt{eclipse-rdf4j/rdf4j} fixed a flaky test and updated the \texttt{AGENTS.md} file to handle flaky tests during test execution.
Analyzing co-changes of AI context files and other artifacts is a promising direction for future work.

In summary, the evolution of the \texttt{AGENTS.md} files in our sample varies, and we did not identify clear patterns in terms of when and how often changes occur.
However, we did identify common change categories.
Based on these categories, it appears that changes are mostly made to fine-tune and adjust instructions. 

\section{Conclusion}

In addition to \texttt{README} files for humans, OSS projects increasingly include AI context files for AI coding agents.
In other words, software developers are now writing and maintaining documentation for machines.
Our results show that conventions for this new software artifact are still in flux.
Projects differ widely in what they encode (e.g., conventions, architecture) and how they express it (e.g., prescriptive vs. prohibitive).
These stylistic variations mirror different prompt writing practices.
Thus, OSS repositories serve as natural laboratories for studying how developers experiment with ``talking'' to agent-based AI tools.

AI context files are maintained software artifacts. They are versioned, reviewed, quality-assured, and tested.
Future work needs to evaluate how their content, structure, and style affect agent behavior and task performance, and how automated feedback loops could update or refine these files based on observed results.
Research should also investigate the co-evolution of source code and related AI context files, similar to the co-evolution of source code and comments~\cite{Fluri2009}.
Open questions include whether standard schemas could improve interoperability, whether repositories should maintain one or multiple AI context files, and how to coordinate instructions for multiple agents.
Beyond technical considerations, this new form of documentation has the potential to reshape communication, review, and collaboration patterns in software teams as instructions move from being written for humans to being negotiated between humans and AI.
The systematic study of AI context files has great potential to provide actionable recommendations to practitioners.

\clearpage

\balance
\bibliographystyle{ACM-Reference-Format}
\bibliography{literature}

\end{sloppy}
\end{document}